# State Tomography of Toroidal Light Pulses

Luka Vignjevic[1], Yijie Shen[2], Nikitas Papasimakis[1] and Nikolay I. Zheludev[1,3]

[1] Optoelectronics Research Centre & Centre for Photonic Metamaterials, University of Southampton, Southampton, UK
[2] Centre for Disruptive Photonic Technologies & The Photonics Institute, Nanyang Technological University, Singapore
[3] Texas A&M University, Institute for Advanced Study, USA

E-mail: l.vignjevic@soton.ac.uk

## Abstract

Toroidal pulses, also known as Focused or Flying Doughnuts (FDs), are few-cycle pulses of doughnut-like topology. Originally proposed by Hellwarth and Nouchi in 1996, they have recently been experimentally realized. Toroidal pulses exhibit intriguing topological features, including skyrmionic field configurations and extensive regions of energy backflow, while their light-matter interactions have been associated with the excitation of toroidal and non-radiating modes in matter. The non-separable nature of toroidal pulses, encompassing both space-time and space-polarization couplings, positions them as promising candidates for robust information and energy transport. However, advancing their fundamental study and deployment in applications requires practical characterization methods, particularly with respect to their non-separability. In this work, we experimentally generate collimated optical toroidal pulses and analyze their space-polarization and space-time couplings using tomographic techniques. We quantify their degree of non-separability through measures such as concurrence and fidelity, benchmarking against ideal FD pulses. The reported results will be of interest to the fundamental study of toroidal pulses and spatiotemporal structured light more broadly, enabling applications in telecommunications, spectroscopy, metrology, and imaging.

# 1. Introduction

Light pulses of toroidal topology, resembling flying or focused doughnuts (FDs), were introduced in 1996 by Hellwarth & Nouchi [1] emerging from a wider family of localised pulses derived by Ziolkowski, known as the modified power spectrum pulses [2]. FDs are exact solutions to Maxwell's equations exhibiting space-time nonseparable structure, meaning the spatial and temporal dependence of the pulse cannot be separated. They were observed for the first time recently in the optical and THz parts of the spectrum [3, 4], followed by demonstration at microwave frequencies [5]. All-optically controlled schemes for the generation of THz FDs were also recently demonstrated [3]. Toroidal pulses complete the field of toroidal electrodynamics as the propagating counterparts of toroidal excitations in matter [6]. They exhibit a fine topological structure with mutliple singuarities and extended areas of energy backflow [7]. Their light-matter interactions are non-trivial and they are predicted to efficiently engage non-radiating charge-current configurations [8]. Recently, FDs were generalized to the family of supertoroidal pulses of which they are the fundamental member. Supertoroidal pulses exhibit intritguing topological properties, including skyrmionic behaviour and self-affine areas of energy backflow [9] that can persist upon propagation over arbitrary distances [10].

The exotic properties of toroidal pulses are intricately linked to their space-polarization and space-time nonseparable nature [11]. Space-polarization nonseparbility is where polarisation depends on spatial position, thus the function describing the spatial amplitude distribution of the light and its polarisation cannot separated into products of eachother as with homogeneously polarised light. Previous work has shown this is similar to a two qubit system describing a particle where entanglement exists between two degrees of freedom [12, 13]. Thus a space-polarization nonseparble state can be expressed as

$$\Psi = \alpha|U_R, x\rangle + \beta e^{i\varphi}|U_L, y\rangle, \qquad (1)$$

where the right and left hand circular polarisation basis states have different spatial modes, thus they can not be factorised into a product of each other. The possible states of this superposition can be represented and visualized by a higher order Poincare sphere [14, 15]. When corellations exist between spatial and temporal frequencies of light, this is known as space-time nonseparability and leads to the spatial separation of the different frequency components of light. Such a coupling between degrees of freedom can result in desireable properties, such as nondiffraction [16]. Due to these properties, nonseparable light could have wide applications in areas such as metrology [17, 18], sensing [19] and communication [20, 21], to name a few.

In the context of toroidal light pulses, this nonseparability manifests in the form of vectorial (radial or azimuthal) polarization and longitudinal field components emerging from the toroidal topology of the pulse. Space-time (or equivalently space-frequency) coupling leads to spatially varying spectral structure, which is known to control the propagation dynamics of pulses [22]. Whereas the fundamental FD is isodiffracting [23], namely the pulse propagates without distortion in its spatio-spectral structure, supertoroidal pulses can be non-diffracting [10]. Thus, quantitavely characterizing the nonseparable properties of toroidal pulses is crucial for the study of their propagation dynamics and light-matter interactions, as well as their deployment in practical applications. However, characterization methods to date are complex and rely on capturing the instantaneous fields with high temporal and spatial resolution [24], which is particularly challenging at high frequencies.

There have been recent devopments where, Digital Micromirror Devices (DMD's) or Spatial Light Modulators (SLM's) are used to make projections onto different spatial modes of the light, performing a modal decomposition [25-29]. When these measurements are made in tandem with polarisation projections, the space polarisation coupling of the light can be fully characterized [30, 31]. However it should be noted that in these past studies the light source has been monochromatic and continuous wave in nature, thus changes are needed to apply these methods to broardband ultrashort pulses. Furthermore, methods have been developed to evaluate the space-time coupling of electromagnetic pulses, where the positions the the inensity peaks for each monochromatic component are measured as a function of propagation distance. This allows the degree of coupling between space and time to be quntified [32].

Here, we apply a state tomography approach to the characterization of space-polarization and space-time nonseparability of experimentally generated toroidal pulses. Using space-polarisation and space-time tomography on the monochromatic components of the generated pulse, we quantify the strength of the nonseparability by concurrence with values of 0.80 and 0.91, in the space-polarization and space-time domains, respectively, while we measure the similarity to the ideal FD pulses by fidelity. Our work provides a practical route to the characterization of nonseparability of toroidal pulses and will be of interest to the fundamental study of structured light pulses and for the applications of toroidal pulses in information and energy transfer.

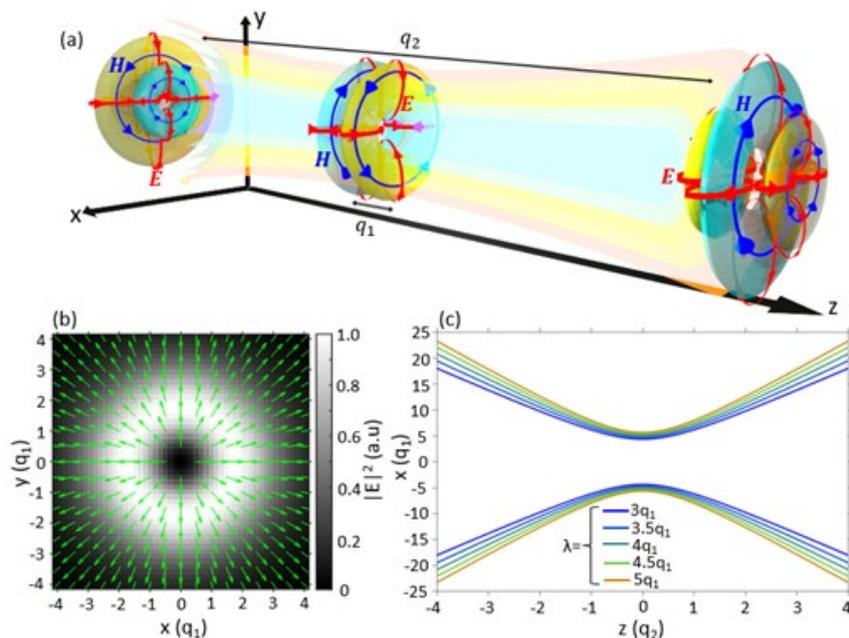

Figure 1. (a) Schematic of a Flying Doughnut pulse as it propagates through the focused region, where $q_1$ is the pulse length (effective wavelength) and $q_2$ is the Rayleigh range. Due to a Guy phase shift, the pulse duration evolves from 1½-cycles away from focus to single-cycle at focus. (b) Intensity (grey-scale image) and polarization (green arrows) profile of the FD in the transverse plane. The spatially varying polarization is a manifestation of the space-polarization no separability in the FD. (c) Evolution upon propagation of the intensity maxima positions for different monochromatic components of the pulse. All monochromatic components of the pulse propagate with the same Rayleigh range, a behaviour known as is diffraction.

## 2. Generation of toroidal light pulses

The electric and magnetic fields of tranverse magnetic (TM) toroidal pulses is given by [1]:

$$E_\rho = -4if_0 \sqrt{\frac{\epsilon_0}{\mu_0}} \frac{\rho(q_2 - q_1 - 2iz)}{[\rho^2 + (q_1 + i\tau)(q_2 - i\sigma)]^3}, \quad (1)$$

$$E_z = 4f_0 \sqrt{\frac{\epsilon_0}{\mu_0}} \frac{\rho^2 - (q_1 + i\tau)(q_2 - i\sigma)}{[\rho^2 + (q_1 + i\tau)(q_2 - i\sigma)]^3}, \quad (2)$$

$$H_\theta = 4if_0 \frac{\rho(q_2 - q_1 - 2iz)}{[\rho^2 + (q_1 + i\tau)(q_2 - i\sigma)]^3}, \quad (3)$$

where $\tau = z - ct$, $\sigma = z + ct$, $q_1$ is a parameter with units of length that plays the role of effective wavelength, $q_2$ defines the depth of the focused region analogous to the Rayleigh range of a Gaussian pulse. Such pulses are radially polarized with longitudinal electric field components (see Figs. 1a&b). The vector nature of the polarization is a form of space-polarisation nonseparability, where the electric field direction varies with the spatial position in the pulse. Transverse electric (TE) toroidal pulses can be derived with an exchange of electric and magnetic fields in Eq. (3), resulting in azimuthally polarization and longitudinal magnetic field components. As the pulse propagates through the focus its structure changes from 1 ½ cycle with 3 lobes to 1 cycle with 2 lobes at focus then back to 1 ½ cycle again with the polarity reversed, a consequence of the Gouy phase shift. Toroidal pulses are space-time nonseparable, whereby the pulse frequency spectrum varies across the transverse plane. This is illustrated schematically in Fig. 1c, where the trajectories of the intensity maxima of different monochromatic components of the pulse are presented. All monochromatic components of the pulse propagate with the same Rayleigh range, resulting in correlations between frequency and angular dispersion, where the higher the frequency, the lower the angular dispersion. Thus, higher frequency components of the pulse are located nearer the center while lower frequencies are further out. These properties result in an effect known as isodiffraction [33], where each frequency component of the pulse retains the same relative position with respect to the other frequencies, rendering the space-time coupling of the pulse propagation invariant.

The experimental setup for the generation of TLP's is outlined in Fig. 2.(a). A Spectra Physics Element laser, produces 10 fs pulses with a FWHM bandwidth of 105 nm, centered at 800 nm which first pass through a pulse compressor, that applies a spectral phase which pre-compensates for the dispersion of the optics further down the beam path. The pulses then passes through a two times beam expander before entering a half waveplate and segmented waveplate, which converts the input linearly polarized pulses into radially polarized ones. Finally the pulses are spatially filtered using a pair of off axis parabolic mirrors and a 100 μm pinhole at the focal plane in between them. Note, in previous experiments where optical FD pulses were generated, a metasurface was placed between the parabolic mirrors instead of a pinhole (see ref. [4] for more details). This metasurface was designed to selectively transmit frequencies which varied with spatial position, thus imparting a specific space frequency coupling. However, it has been found that for collimated pulses, a 100 µm pinhole is able to recreate the same space frequency coupling. An image of the experimentally generated pulse is shown in Fig. 2b. The spectrum of the experimentally generated pulses, with the coresponding spectral phase, measured using an FC SPIDER (Spectral Phase Interferometry for Direct Electric field Reconstruction) made by APE is given in Fig. 2.c. The FWHM of spectrum was found to be 90nm, centred at around 805nm and the spectral phase was found to be relatively flat with phase fluctuations under 1 radian over a 150 nm wavelength range. This corresponds closely to the ideal FD pulse, which has a flat spectral phase and indicates the pulse is compressed close to the Fourier limit. Time domain measurements were also collected using the SPIDER and presented in Fig. 2d, showing the pulse envelope and corresponding phase. The pulses have a FWHM duration of 15 fs, which equates to approximately 6 optical cycles. In contrast to our earlier work [4], the generation of toroidal pulses here does not employ a metasurface element following recent observations of the "self-healing" propagation dynamics of imperfect toroidal pulses [5].

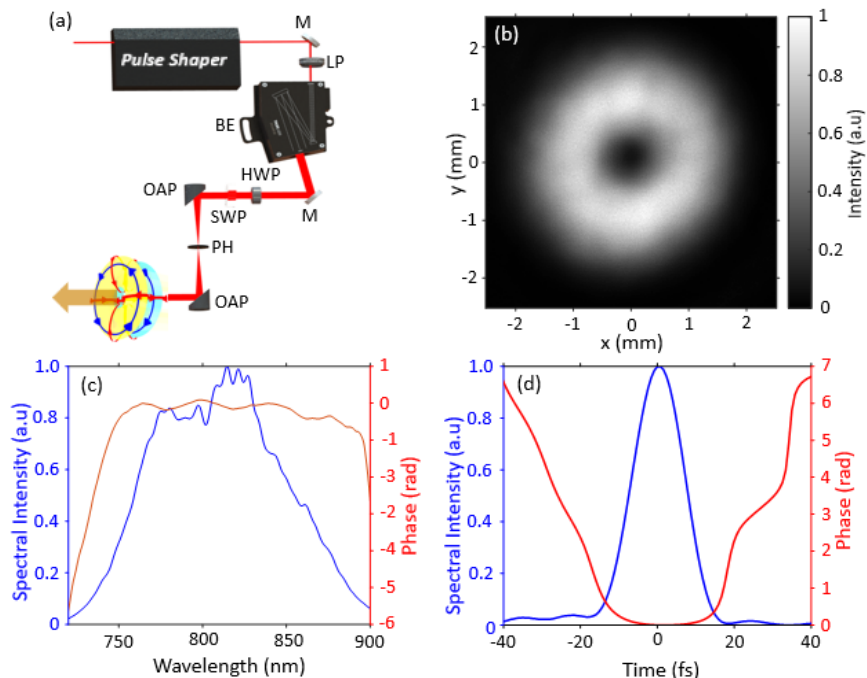

Figure 2. (a) Experimental setup for the generation of toroidal pulses. A 10 fs pulse from a Ti:Sa laser passes through a pulse shaper and a 2x beam expander. A half wave plate selects the polarization that enters the segmented waveplate, which converts linear polarization to radially or azimuthal. The vector polarized pulse is then passed through a spatial filter that removes scattering related artefacts and collimates the pulse. (b) Intensity profile of the generated pulse after collimation. (c-d) Spectral and temporal forms of the generated pulse as obtained by SPIDER measurements.

# 3. Space-polarization tomography

To quantify the space-polarization nonseparability of the generated toroidal pulses, we follow a state tomography approach typically employed in the diagnostics of vectorial light beams [29]. The approach is analogous to Stokes polarimetry, except projections are taken in the spatial and polarization degrees of freedom, extending the state space from two to four dimensions. The projective measurements are used to calculate the density matrix, which encapsulates all the space-polarisation information about the light:

$$\rho = \left(\frac{1}{2}\sum_{m=0}^{3}\rho_m\sigma_m\right)^S \otimes \left(\frac{1}{2}\sum_{n=0}^{3}\rho_n\sigma_n\right)^P = \frac{1}{4}\sum_{m,n=0}^{3}\rho_{mn}\sigma_m^S\otimes\sigma_n^P \qquad (4)$$

where $\rho_{mn}$ is a $4 \times 4$ matrix of coefficients (please see the suplementary material for more information), the higher dimensional analogue of the Stokes vector, and $\sigma_m, \sigma_n$ are the Pauli spin matrices. This is the Kronecker product of density matrices characterising the spatial and polarization states of the light, as denoted by the $S$ and $P$ superscripts respectively. This results in a four-dimensional space Hilbert space on a higher dimensional Poincare sphere.

The state tomography approach is applied to each monochromatic components of the toroidal pulse. These can be decomposed into a superposition of left and right hand circularly polarised beams with Orbital Angular Momentum (OAM) values of $+1$ and $-1$ respectively. This can be expressed using bra-ket notation as $\psi_{CV} = cos\theta \cdot |l=1\rangle|R\rangle + e^{i\phi} \cdot sin\theta \cdot |l=-1\rangle|L\rangle$, where varying the phase between the two terms switches the resultant between radial and azimuthal polarisations. This subset of the vector beams can be represented by a Poincare like sphere where the azimuthal, spiral and radial polarisation states exist along the equator, with spin-orbit coupled beams at the poles.

The experimental setup for space-polarizaton tomography follows ref. [34] with the addition of a bandpass filter that selects a monochromatic component of the toroidal pulse (see Fig. 3a). a quarter waveplate, half waveplate and linear polarizer make polarization projections (see Supplementary). Note, the use of a half wave plate means all polarization projections are rotated into the horizontal plane, mitigating any errors caused by polarization dependence of any subsequent optical components. The light then incidents on a Digital Micromirror Device (DMD) with spatial mode phase masks encoded which makes projections onto spatial mode eigenstates. Example of spatial phase masks can be seen in Fig. 3b. The resulting diffraction pattern is then imaged in the Fourier plane of a lens. Examples of numerically calculated diffraction patterns corresponding to an ideal FD pulse are shown in Fig. 3c. The projection value is given by the on-axis intensity of the first diffraction order. To obtain the relative phases between these eigenstates, an overcomplete set of measurements is required, consisting of all the permutations of 6 spatial and 6 polarization projections (36 tandem projections).

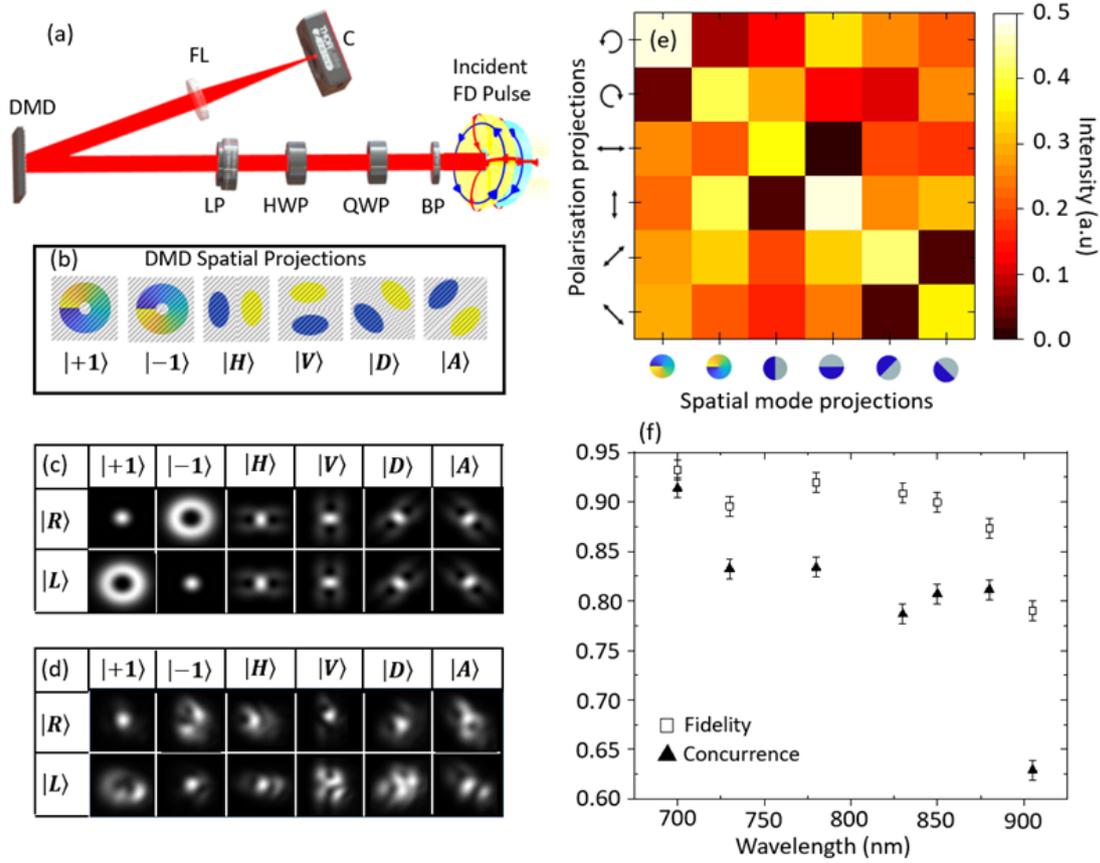

Figure 3. (a) Schematic of the experimental setup used for space-polarisation tomography of toroidal pulses. A bandpass filter (BP) is used to isolate a monochromatic component of the pulse, a quarter waveplate (QWP), half waveplate (HWP) and linear polariser (LP) are used to make polarisation projections. Spatial mode projections are made by a Digital Micromirror Device (DMD), the resulting diffraction orders are projected into the far field by a Fourier lens (FL) before being capture by the camera (c). (b) DMD masks and their corresponding spatial modes. (c-d) Images of theoretically calculated (c) and experimentally recorded (d) first order diffraction patterns of toroidal pulse incident on the DMD. (e) Matrix of tandem spatial mode and polarization projective measurement values at. (f) Fidelity and concurrence values as a function of wavelength.

Space polarization tomography of toroidal pulses was performed by selecting seven different monochromatic components at wavelengths, λ, of 700 nm, 730 nm, 780 nm, 830 nm, 850 nm, 880 nm, and 905 nm, covering the full extent of the pulse bandwidth. Characteristic diffraction patterns are presented in Fig. 3d for λ=700 nm. From the on-axis intensity of different diffraction patterns, a tomography matrix can be constructed for each monochromatic component of the pulse. An example of such a matrix is presented in Fig. 3e for λ=780 nm, indicating that this pulse component can be decomposed into the superposition of an RCP beam with +1 OAM and an LCP beam with -1 OAM, which is consistent with a radially polarized (TM) pulse. Similar results are obtained for all wavelengths (see Supplementary). The space-polarization nonseparability is quantified by the concurrence, C, which take values between 0 and 1, corresponding to homogeneous, separable, light and maximally nonseparable light, respectively. The concurrence for the experimentally characterized FD is plotted in Fig. 3f as a function of wavelength. Concurrence is high across the pulse bandwidth with an average value of <C>=0.80 indicating a strong coupling between space and polarisation degrees of freedom. The likeness of the experimentally generated pulse to the theoretical FD is quantified by fidelity, F, which varies from 0 to unity. The average fidelity is <F>=0.88, indicating the space polarization structure of the pulse closely resembles the radially polarized nature of the ideal TM FD. Further, the mirrors used have differences in reflectivity between *s* and *p* polarizations, which increases with increasing wavelength within the bandwidth under consideration, causing the space polarization state of the vector pulse to become distorted. This potentially explains the general decrease observed in the fidellity and concurrence. From the space-polarization tomographic measurements, we can reconstruct the polarization state of the toroidal pulse. A characteristic example is shown in Fig. 4a for λ=780 nm, indicating that this component is predominantly linearly polarized with the electric field along the radial direction. Similar results are obtained for all monochromatic components of the toroidal pulse (see Fig. 4b).

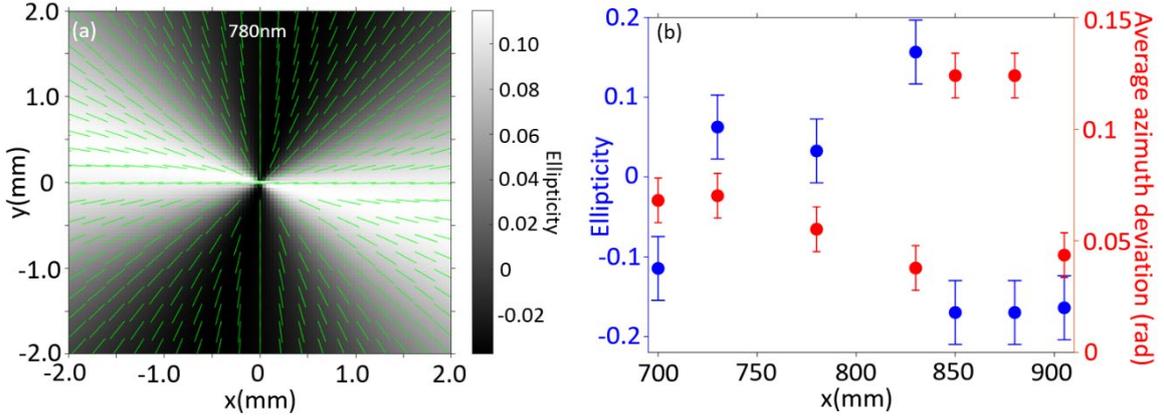

Figure 4. Space-polarization nonseparability in generated TM toroidal pulses. (a) Ellipticity (gray-scale image) and azimuth rotation (green arrows) at λ=780nm. (b) Average ellipticity and azimuth deviation from the radial direction as a function of wavelength. The average is taken over the transverse plane.

## 4. Space-time tomography

Space-time (or space-frequency) tomography was performed by taking the angular spectra of monochromatic components of the pulse, using a Fourier lens and a camera, placed after the bandpass filter and before the quarter waveplate (see Supplementary S5). This allowed the angular divergence of each monochromatic component to be calculated and compared with theoretical predictions for the ideal TLP allowing the $q_2$ value of the experimentally generated pulse to be calculated. The relationship between angular divergence and Rayleigh range is given by the following expression

$$Z_r = \frac{\lambda}{\Theta_{div}^2 \pi}, \qquad (5)$$

where $Z_r$ is the Rayleigh range, $\Theta_{div}$ is the angular divergence and $\lambda$ is the wavelength. Using this information, a density matrix is calculated for the space and time degrees of freedom, where wavelength and angular divergence are descretised into five separate states each. From this matrix, the similarity of the space-time coupling to the ideal TLP can be quantified in the form of the fidelity and the degree of nonseparability in the form of the concurence [32].

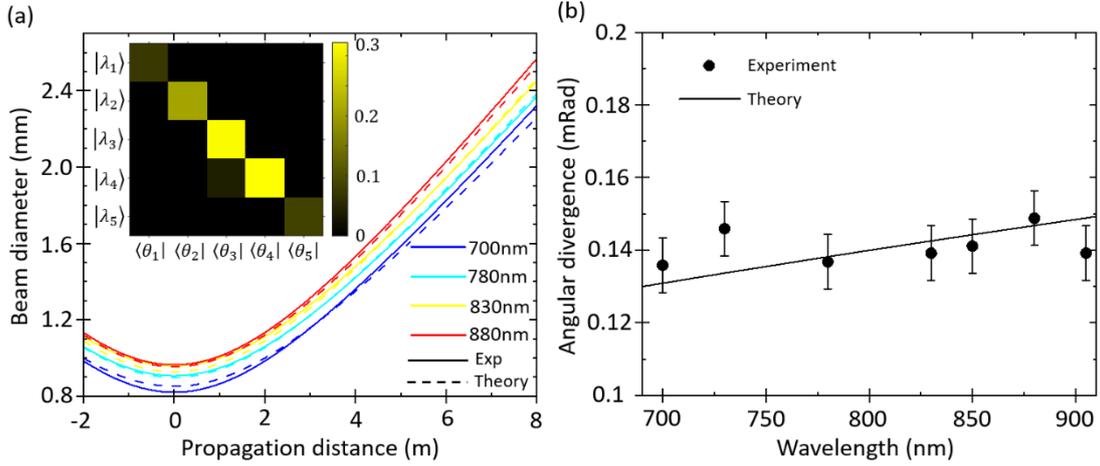

Figure 5. (a) Trajectories of the intensity maxima of monochromatic components of the experimentally characterized (solid line) and ideal (dashed line) toroidal pulse. Inset: Space-frequency tomography matrix showing the correlation between frequency and angular divergence, where $|\lambda_1\rangle, |\lambda_2\rangle, |\lambda_3\rangle, |\lambda_4\rangle, |\lambda_5\rangle$ correspond to wavelengths of 700, 780, 830, 850, and 905 nm respectively. $\langle\theta_1|, \langle\theta_2|, \langle\theta_3|, \langle\theta_4|, \langle\theta_5|$ correspond to angular divergences of 0.131, 0.138, 0.143, 0.144, 0.149 mrad, respectively. (b) Experimentally measured (circles) and theoretical (solid line) angular divergence of the toroidal pulse components as a function of wavelength.

The results of space-time tomography of the generated toroidal pulses are shown in Fig. 5a in the form of trajectories of the intensity maxima of the pulse monochromatic components. The trajectories do not cross each other indicating that the pulse is isodiffracting and follow closely to the theoretical ones (corresponding to $q_1 = 200$nm and $q_2 = 1.68 \times 10^7 q_1$). The corresponding tomography matrix is show in the inset to Fig. 5a, from which we calculated the corresponding fidelity to be 0.72 and concurrence of 0.91 indicating that the space-time (or equivalently space-frequency) profile of the pulse closely resembles that of an ideal FD and is highly noseparable. This illustrated by the angular divergence of each monochromatic component (see Fig. 5b), where all but one of the experimental and theoretical values deviate by less than 0.01 mrad. Such deviations from the space-time profile of the ideal FD are attributed to the fact that the generation scheme employed here did not make use of a gradient metasurface (as for instance in ref. [4]).

## 5. Conclusions

In summary, we have characterized quantitatively the nonseparable properties of experimentally generated toroidal light pulses. In particular, we have applied state tomography to quantify the space-polarization and space-time coupling of toroidal pulses, which define their topological properties and propagation dynamics. Importantly, we have showed that even in the absence of a gradient metasurface element, ultrafast laser pulses evolve towards the ideal FD pulses in terms of their space-time nonseparability, in accordance with recent works [5]. Our results pave the way towards higher-dimensional tomography of spatiotemporally structured light and will facilitate the deployment of nonseparable broadband beams and pulses in telecommunications, spectroscopy, and metrology applications.

## Acknowledgements

The authors acknowledge the support of the European Research Council (advanced grant FLEET-786851, Funder Id: https://doi.org/10.13039/501100000781).

## Data availability

The data from this paper can be obtained from the University of Southampton ePrints research repository https://doi.org/10.5258/SOTON/DXXX.